# The prototype string for the km$^3$-scale Baikal neutrino telescope

V.Aynutdinov,[a,*] A.Avrorin,[a] V.Balkanov,[a] I.Belolaptikov,[d] D.Bogorodsky,[b] N.Budnev,[b] I.Danilchenko,[a] G.Domogatsky,[a] A.Doroshenko,[a] A.Dyachok,[b] Zh.-A. Dzhilkibaev,[a] S. Fialkovsky,[f] O. Gaponenko,[a] K.Golubkov,[d] O. Gress,[b] T. Gress,[b] O.Grishin,[b] A.Klabukov,[a] A. Klimov,[h] A.Kochanov,[b] K.Konischev,[d] A.Koshechkin,[a] V.Kulepov,[f] D.Kuleshov,[a] L.Kuzmichev,[c] E.Middell,[e] S.Mikheyev,[a] M.Milenin,[f] R.Mirgazov,[b] E Osipova,[c] G.Pan'kov,[b] L.Pan'kov,[b] A.Panfilov,[a] D.Petukhov,[a] E.Pliskovsky,[d] P.Pokhil,[a] V.Poleschuk,[a] E.Popova,[c] V.Prosin,[c] M. Rozanov,[g] V.Rubtzov,[b] A.Sheifler,[a] A.Shirokov,[c] B.Shoibonov,[d] Ch.Spiering,[e] O.Suvorova,[a] B.Tarashansky,[b] R.Wischnewski,[e] I.Yashin,[c] V.Zhukov[a]

[a] *Institute for Nuclear Research, 60th October Anniversary pr. 7a, Moscow 117312, Russia*

[b] *Irkutsk State University, Irkutsk, Russia*

[c] *Skobeltsyn Institute of Nuclear Physics MSU, Moscow, Russia*

[d] *Joint Institute for Nuclear Research, Dubna, Russia*

[e] *DESY, Zeuthen, Germany*

[f] *Nizhni Novgorod State Technical University, Nizhni Novgorod, Russia*

[g] *St.Petersburg State Marine University, St.Petersburg, Russia*

[h] *Kurchatov Institute, Moscow, Russia*

**Abstract**

A prototype string for the future km$^3$-scale Baikal neutrino telescope has been deployed in April, 2008 and is fully integrated into the NT200+ telescope. All basic string elements – optical modules (with 12"/13" hemispherical photomultipliers), 200MHz FADC readout and calibration system – have been redesigned following experience with NT200+. First results of in-situ operation of this prototype string are presented.

---

[*] Corresponding author. 8-495-123-35-08; fax: +7-499-783-92-98; e-mail: aynutdin@pcbai10.inr.ruhep.ru

## 1. Introduction

The Baikal Neutrino Telescope is operated in Lake Baikal, Siberia, at a depth of 1.1 km. The first stage telescope configuration NT200 [1,6] started full operation in spring 1998. NT200 consists of 8 strings, each with 24 pairwise arranged optical modules (OM). Each OM contains a 37-cm diameter hybrid photodetector QUASAR-370, developed specially for this project [2]. The upgraded Baikal telescope NT200+ [3] was commissioned in April, 2005, and is made of a central part (the old, densely instrumented NT200 telescope) and three additional external strings, as displayed in Fig.1

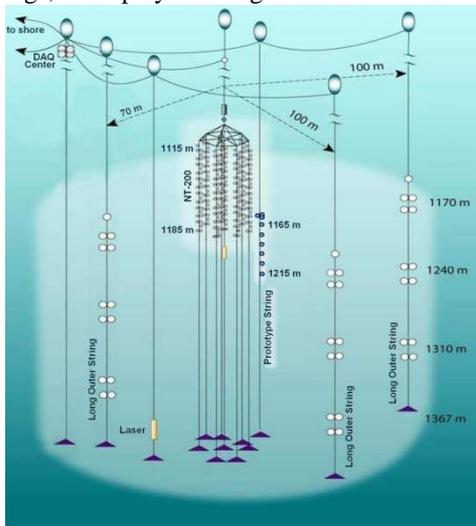

*Fig. 1. The Baikal Telescope NT200+ and the prototype string.*

The construction of NT200+ was a first step towards a $km^3$-scale Baikal neutrino telescope. Such a detector could be made of building blocks similar to NT200+, but with NT200 replaced by a single string, still allowing separation of high energy neutrino cascades from background. It will contain a total of 1300-1700 OMs, arranged on 90-100 strings with 12-16 OMs each, and an instrumented length of 300-350 m. Interstring distances will be ≈100 m. The effective volume for cascades with energy above 100 TeV is 0.5-0.8 $km^3$, the threshold for muons is in the range 10-30 TeV.

The existing NT200+ allows to verify key elements and design principles of the $km^3$ (Gigaton-volume) Baikal telescope. The next milestone of the ongoing research and development work for the $km^3$ telescope was the installation of a "new technology" prototype string as part of NT200+ (Fig. 1).

The basic goals of the prototype string installation are: (1) investigation and in-situ test of basic elements of the future detector (new optical modules, DAQ system and cabling system), (2) studies of the basic DAQ/triggering approach for the $km^3$-detector, and (3) comparison of the classical TDC/ADC approach with a FADC-based full pulse shape readout.

## 2. Prototype string design

The design of the prototype string is presented in Fig. 2.

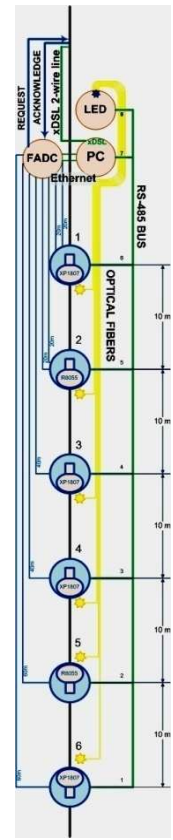

*Fig. 2. The $km^3$-prototype string*

The string consists of 6 new generation optical modules with photomultipliers R8055 (Hamamatsu) and XP1807 (Photonis). The distance between OMs along the string is 10 m. The upper 4 OMs have photomultipliers with downward looking photocathode, while the two bottom OMs contain PMs with upward looking photocathode. The preamplified dynode outputs of all 6 PMs are connected through underwater coaxial cables to 200 MHz FADC boards, located in a separate glass sphere. Two FADC channels are used to capture the waveforms of 2 additional low-gain channels of the two upper PMs.

Data from the FADC unit are transmitted via a local Ethernet line to the underwater micro-PC unit. Synchronization of prototype string and telescope data acquisition systems is the same as for the outer strings of NT200+ [3]. Time and amplitude calibration is provided by a string LED flasher located in a separate glass sphere near the FADC and PC units. Light pulses from the flasher are transmitted to each OM via optical fibers with calibrated length. Control and monitoring of OM and LED flasher operation is provided by the string PC unit via a RS-485 underwater bus.

### 2.1. The new Optical Module

The new optical module is a key element of the future km3-scale neutrino telescope. The block diagram of the OM is presented in Fig. 3. It consists of a pressure-resistant glass sphere with 42 cm diameter, housing the photomultiplier with its hemispherical photocathode embedded in optical gel. A high permittivity alloy cage surrounds the tube, shielding it against the Earth's magnetic field.

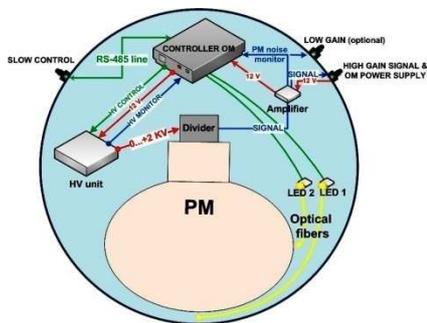

*Fig. 3. The new generation Baikal optical module, with a large-area hemispherical photomultiplier*

Analog PM signals and slow control data are transmitted through two pressure-resistant coaxial connectors (optionally three for low-gain readout). Signal cables are used also for the 12 VDC power supply of the OM.

Photomultipliers of two types are used as light sensors for the OMs: Photonis-XP1807 with 12" photocathode diameter and Hamamatsu-R8055 with a 13" photocathode. Preliminary tests have shown that these tubes have approximately the same response to Cherenkov light [4].

The OM electronics consists of a High Voltage (HV) unit, a two channel preamplifier and a controller. The HV unit provides the power for the tube divider in the range from 0 up to 2 kV. The tube gains were adjusted to about $10^7$, which corresponds to a HV between 1250 and 1650 V for the different tubes. An additional signal amplification is provided by a two channel fast preamplifier. Single photoelectron (pe) pulse amplitudes are about 25 mV for the high gain channel, and about 3 times smaller for the low gain channel. This corresponds to a spectrometric channel linearity range up to about 500 pe. Only the two upper OMs of the string are operated in the two-gain operation mode. The other OMs operate with only the high-gain output. For these OMs, the second amplifier output is used for PM noise monitoring with the OM controller.

The OM controller is designed on the basis of the microcontroller SiLabs C8051F121. The unit is intended for HV value regulation and monitoring, for permanent PM noise measurement, for monitoring of the low voltage power supply and the temperature inside the glass sphere, and for calibration of the spectrometric channel with LED pulses. Two independent LED channels are foreseen for calibration. The dominant wavelength of the LED is 445 nm and the pulse has a width of ~5ns FWHM. The possibility of independent regulation of the LED light intensity and low cross talk between LED channels (<1%) allow to directly test the linearity of the spectrometric channel. Another important LED flasher feature is the possibility to change the delay between flashes from 0 up to 1000 ns.

Slow control data to and from the OMs are transferred under the control of the string PC unit via an underwater RS-485 bus, using a single coaxial cable.

## 2.2. The FADC underwater unit

An important feature of the prototype string is the possibility to measure the pulse shape for each PM channel. PM signals arrive at the FADC unit, where they are digitized by custom-made 4-channel FADC VME modules with 200 MHz sampling rate [5]. The underwater FADC unit is shown in Fig. 4.

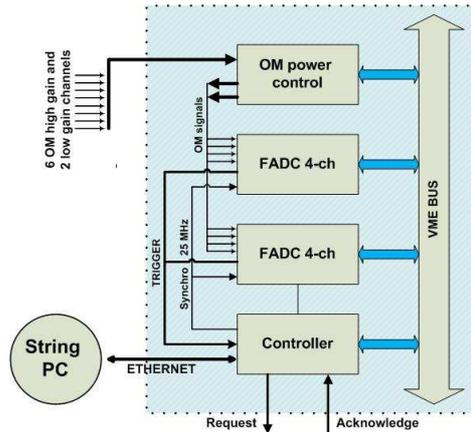

*Fig. 4. The underwater FADC unit*

This unit consists of two FADC modules, an OM power controller and a VME controller. The VME controller provides trigger logic, data readout from FADC boards and connection via local Ethernet to the string underwater PC. The trigger is formed by a coincidence of $\geq$ n OMs pulses above threshold within a selectable time window.

Every FADC board consists of four simultaneously digitizing ADCs on the basis of AD9430 microcircuits. Every FADC has 12 bit resolution and samples at 200 MHz. Two channels of the FADC board are shown in Fig. 5.

The digitized signals from each ADC are transferred to a FPGA which handles the data. A double-buffered FIFO memory of 1024 counts depth minimises the readout dead-time: while one buffer is ready for VME-readout, the second is connected to the ADC output. Receiving a trigger signal from the master device, 512 time slots before and 512 after the trigger are captured. The high gain OM-channel is connected to an adjustable digital comparator, which works out the request to the trigger logic, located on the controller. This signal is also connected to an amplitude analyzer, which accumulates monitor histograms.

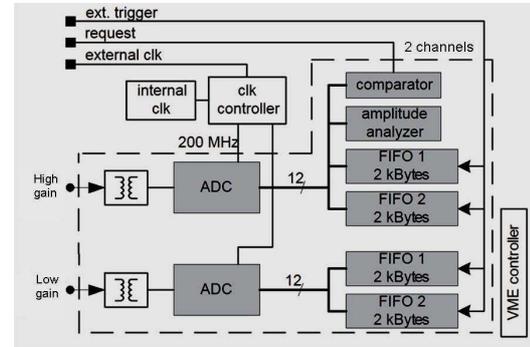

*Fig 5. Internal structure of the FADC board*

The power control board (Fig. 4) is intended for the OM power supply. This module separates input analog pulses and +12 VDC OM power, and allows to switch power on/off for each optical module independently.

## 2.3. The String PC unit

Data from the FADC unit are transmitted via a local Ethernet connection to the underwater micro-PC, located inside the String PC unit (Fig. 6), for on-line analysis and data compression.

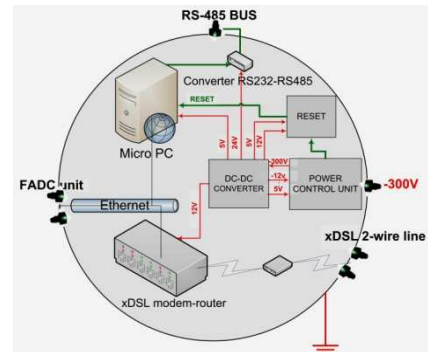

*Fig. 6. The String PC unit*

Communication between the string PC-unit and the underwater control centre of NT200+ ([4], Fig. 1) is provided by 2-Mbit DSL modems (FlexDSL) via a 2-wire line of ~1 km length. Slow control of OMs and the LED flasher is provided by the string PC unit via a RS-485 underwater bus. The main slow control functions are regulation of the PM high voltage, control of the LED flasher intensity and pulse delay, and measurement of the PMT noise rates.

Using a separate underwater PC unit is a temporary solution. This unit allows flexibility while optimizing the data acquisition and on-line underwater data analysis. In future, the string-PC functionality will be integrated into the FADC unit.

*2.4. The LED flasher unit*

Time and amplitude calibration are provided by a string LED flasher unit, shown in Fig. 7. The unit is located in a separate glass sphere close to the FADC and PC units.

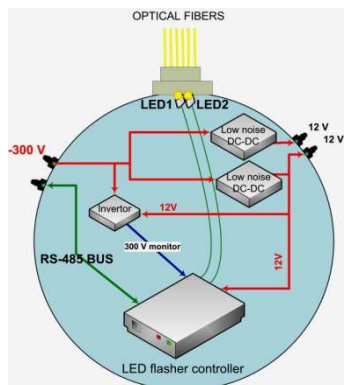

*Fig. 7. LED flasher unit*

Basic elements of the LED flasher unit are two LED drivers and a LED controller. The LED controller has the same design as the OM controller, and provides the possibility of LED pulse amplitude and delay regulation independently for the two LEDs. It also monitors the 300 VDC string power supply.

Light pulses from the flasher are transmitted to each OM via individual optical fibers (as for all NT200+ OMs [1] [4]). The LED flasher allows to measure relative time shifts of the channels and provides the possibility to monitor the single electron spectrum for all PMs of the string.

The LED flasher glass sphere also houses two low noise DC-DC converters (300 V to 12 V) for the OM power supply: ripple & noise (max.) is about 7 mV p-p (significantly below the OM single photoelectron amplitude). This converter was specially designed for the NT200 telescope and has been successfully used to the present time.

## 3. Prototype string deployment and integration with NT200+

The prototype string was installed as a part of NT200+ in April 2008. It is located at a radial distance of ~64 m from the NT200 heptagon center, between the 1-st and 3-rd NT+ external strings (Fig.1). The upper optical module of the new string is deployed at the 1165 m depth, slightly above the lowest NT200 OMs. This prototype string location provides good conditions for cascade detection jointly with NT200. The upward orientation of the two bottom OMs of the string increases the efficiency of cascade event detection.

Data transfer and time synchronization for the prototype string and NT200+ are identical (Fig 8).

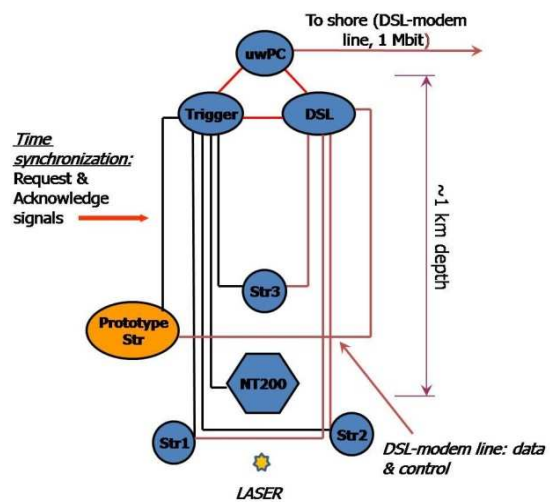

*Fig. 8. Data acquisition systems for the prototype string and NT200+*

The interstring time synchronization is done by measuring the time interval between request pulses produced by an external string and the prototype string with the Central Trigger unit [4]. This unit also produces a common acknowledge signal that initiates the data readouts via the string controllers.

For the connection between the prototype string and the underwater control centre of NT200+ an armoured 1 km carrier cable is used (custom designed for this new setup). This cable includes 2 coaxial lines for request and acknowledge signals, a screened twisted pair for the DSL connection and 2 power lines.

A wire rope of 270 m length is connected to the bottom of the carrier cable. All glass spheres with electronics and OMS and all inter-sphere cables were attached to the wire rope. An OM installed on the rope, is shown in Fig. 9.

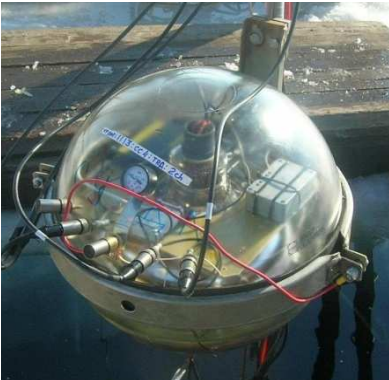

*Fig. 9. An OM with phototube XP1807 after final mounting*

The FADC, String PC, and LED flasher spheres were mounted to the rope 5 m above the uppermost OM, see Fig. 10.

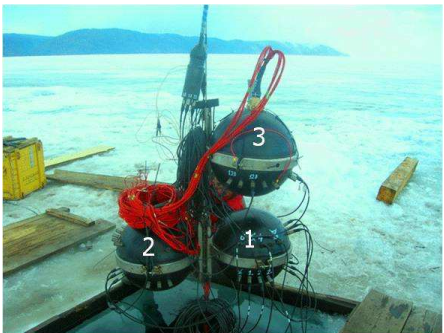

*Fig. 10. FADC sphere (1), String PC unit (2), and LED flasher sphere (3) after final installation*

The underwater connections between glass spheres are based on one-wire shielded coaxial cables and connectors like for the NT200 cable system. In particular, the 100 Mbit local Ethernet line between the FADC and PC units consists of 2 such coaxial cables with signals transferred via wire and screen. In-situ experiments showed, that the contact of the cable screen with the Baikal water (on the connector bulkhead) does not influence the data connection quality. The same approach was also used for the design of the underwater RS-485 cable line.

The first experience with mounting the prototype string shows, that whole deployment time is about 5 hours. This time includes also in-situ OM functionality tests after mounting each optical module on the rope.

## 4. First experimental results

The prototype string has been taking data since April 2008. Two basic modes of string operation are available: joint operation with NT200+ and standalone operation. A coincidence of a prototype string trigger with an external string trigger is necessary in the first mode of operation. This mode is used for investigations of cascade and muon events detected jointly with NT200+ and the new string, while the standalone mode is used for in-depth performance and stability verification with atmospheric muons and light calibration sources.

In the second mode, a 3-fold coincidence of OMs of the prototype string is used as a trigger. Below, we present some preliminary results of tests of the string response to LED flasher and Laser calibration sources in the standalone mode.

The first step of data analysis was a study of the accuracy of the pulse time measurement with the 200 MHz FADC. For this purpose, delayed pulses produced by the LED flasher were applied. An example of such a LED flasher event for the 3$^{rd}$ FADC channel is given in Fig. 11.

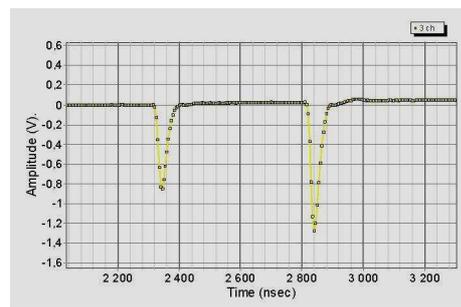

*Fig. 11. Example of a two-pulse LED flasher event*

The first pulse is produced by LED1, the second by LED2 (see Fig. 3). A pulse delay of 497 ns was installed by the LED flasher controller. For these two pulses, detection times and the corresponding delay were calculated from the FADC data. Two different approaches were used for the time calculation: a fit of the pulse leading edge and a full pulse shape fit, respectively. The measured delay time values are given in Table 1 for all FADC channels. The second and third columns indicate average delay and its RMS obtained by the pulse leading edge (LE) fit, column 4 and 5 show the pulse shape fit (PS). The last column presents the number of photoelectrons for the smallest LED pulse. We find the average delay values to be close to the nominal value of 497 ns. Average RMS for two methods are 2.8 ns and 1.8 ns, respectively. The pulse shape approximation gives somewhat better results, but cannot be applied for all types of pulses. For example, pulses of the $1^{st}$ and $4^{th}$ channels here are out of FADC range, and pulse shape fitting is impossible (see rows 1 and 4 in Table 1). An example of a delay distribution is presented in Fig. 12 for the $3^{rd}$ FADC channel.

Table 1.

Time accuracy estimation for the prototype string FADC system

| Channel | Delay, ns (LE) | RMS, ns (LE) | Delay, ns (PS) | RMS, ns (PS) | Qmin, pe |
|---|---|---|---|---|---|
| 1 | 498,50 | 2,70 | - | - | >100 |
| 2 | 497,20 | 2,63 | 497,45 | 1,39 | 19 |
| 3 | 497,05 | 1,41 | 497,15 | 0,79 | 100 |
| 4 | 496,90 | 2,424 | - | - | >100 |
| 5 | 497,15 | 4,60 | 496,80 | 2,97 | 10 |
| 6 | 498,40 | 3,66 | 497,00 | 2,57 | 10 |
| 7 | 496,80 | 1,88 | 497,20 | 1,06 | 75 |

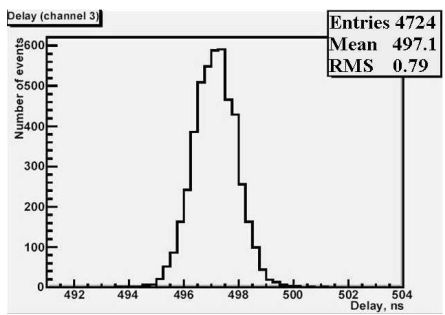

*Fig. 12. Pulse delay distribution for $3^{rd}$ FADC channel (see Table 1)*

A basic function of the LED flasher is the direct measurement of the relative time shifts of the spectrometric channels. Time shifts were determined as the time difference between pulses on different FADC channels for simultaneous detected LED flashes. An example of a LED flasher event for this mode of operation is presented in Fig. 13.

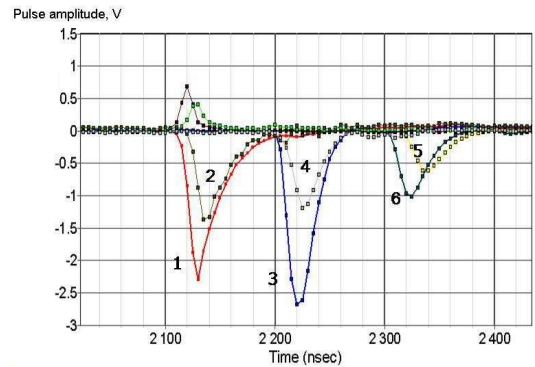

*Fig. 13. Example of a LED flasher event detected with all FADC channels*

In Fig. 13, the indices indicate the OMs; positive pulses correspond to the low gain channels of the $1^{st}$ and $2^{nd}$ OM. Significant time differences between the pulses are caused mainly by differences in signal cable lengths, which are equal to about 20 meters for successive OM pairs. The time shifts were calculated taking into account the relative light pulse delay in the fiber optic cables for all channels.

The LED flasher also allows a direct amplitude calibration of the OM spectrometric channels. Two pulses with a delay of about 500 ns are produced by the LED flashers during the calibration. The first LED provides a small light pulse (single electron mode of phototube operation). The pulse of the second LED has a value significantly larger than the PM noise amplitude. This pulse is used as a trigger for the phototube dark noise suppression. Single electron spectra (SES) were in-situ measured for all OM phototubes. An example of an SES for the $1^{st}$ OM (a XP1807) is presented in Fig. 14. The single electron distribution has an average value of about 80 FADC channels after pedestal subtraction (see parameter p3 in Fig. 14). This value is used as amplitude calibration coefficient for the given FADC channel.

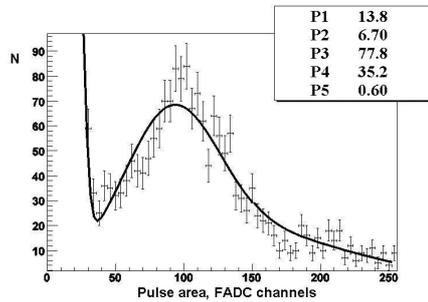

*Fig. 14. Single electron spectrum measured for the 1st OM during in-situ calibration procedure*

The results of the time shift and amplitude calibration were used for the analysis of the prototype string calibration with the NT200+ Laser [4], which is located at a horizontal distance of ~100 m from the string, and at ~1280 m depth (Fig. 1). An example laser event, after time and amplitude calibration, is presented in Fig. 15 (numbers indicate OMs).

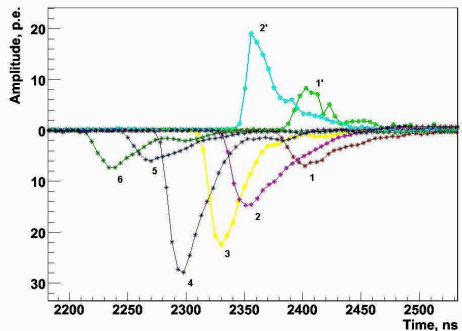

*Fig. 15. Example of a Laser event*

From a preliminary analysis of the laser data we conclude that the pulse arrival times are in good agreement with the expected values (~30ns time difference between neighbouring OMs). A more detailed analysis, allowing to compare parameters of PMs of different types, is in progress now.

## 5. Conclusion

The goal of the Baikal Project is the creation of a $km^3$-scale (gigaton-volume) neutrino telescope in Lake Baikal. The construction of NT200+ in spring 2005 - a detector with about 5 Mton enclosed volume - was a first step towards a $km^3$-scale Baikal neutrino telescope. NT200+ allows to verify key elements and design principles of the km3 detector.

For detailed investigations and in-situ tests of new basic elements of the future detector (new optical modules, DAQ system, cable communications), a "new technology" prototype string with six large area hemispherical PMs and with FADC readout technology was designed and installed in spring 2008 in Lake Baikal. The data acquisition system of the prototype string was fully integrated into the NT200+ telescope.

The prototype string is successfully operating now. First in-situ tests of the prototype string with the underwater laser, the LED flasher and atmospheric muons show good performance of all string elements.

## Acknowledgments

This work was supported by the Russian Ministry of Education and Science, the German Ministry of Education and Research and the Russian Fund of Basic Research (grants 08-02-00432, 08-02-10010, 08-02-00198 and 08-02-10001), and by the Grant of President of Russia NSh-4580.2006.2., and by NATO-Grant NIG-9811707(2005).

## References

[1] I.Belolaptikov et al., Astropart. Phys. 7, 263 (1997)
[2] R.Bagduev et al., Nucl. Inst. Meth. A420, 138 (1999)
[3] V.Aynutdinov et al., Nucl. Inst. Meth. A567, 433 (2006)
[4] K.Antipin et al., Proc. of the 30th ICRC Conference, Merida, Mexico (2007), paper-1084, arXiv:0710.3063
[5] N.Budnev et al., Proc. of the 10th ICATPP Conference, Como, Italy (2007), arXiv:0804.0856
[6] V.Aynutdinov et al, this confernce: " The BAIKAL neutrino experiment: status, selected physics results, and perspectives"